\documentclass{aastex62}
\usepackage[utf8]{inputenc}
\usepackage{amsmath,amsfonts,amssymb}
\usepackage{subfigure}
\usepackage{multirow}
\usepackage{longtable,threeparttablex}
\usepackage{epstopdf}
\usepackage{graphicx,enumerate}
\usepackage{hyperref}
\usepackage{enumitem}
\usepackage{natbib}
\usepackage{xcolor}
\shorttitle{Radio-Discovered Tidal Disruption Event CNSS J0019+00}
\shortauthors{Anderson et al.}

\begin{document}
\title{Caltech--NRAO Stripe 82 Survey (CNSS). III. The First Radio-discovered Tidal Disruption Event, CNSS J0019+00}

\author{M. M.~Anderson}
\affiliation{California Institute of Technology, 1200 E California Blvd. MC 249-17, Pasadena, CA 91125, USA}

\author{K. P.~Mooley}
\affiliation{National Radio Astronomy Observatory, P.O.\,Box O, Socorro, NM 87801, USA}
\affiliation{California Institute of Technology, 1200 E California Blvd. MC 249-17, Pasadena, CA 91125, USA}

\author{G.~Hallinan}
\affiliation{California Institute of Technology, 1200 E California Blvd. MC 249-17, Pasadena, CA 91125, USA}

\author{D.~Dong}
\affiliation{California Institute of Technology, 1200 E California Blvd. MC 249-17, Pasadena, CA 91125, USA}

\author{E. S.~Phinney}
\affiliation{Theoretical Astrophysics, MC 350-17, California Institute of Technology, Pasadena, CA 91125, USA}

\author{A.~Horesh}
\affiliation{Racah Institute of Physics, The Hebrew University of Jerusalem, Jerusalem, 91904, Israel}

\author{S.~Bourke}
\affiliation{Department of Space, Earth and Environment, Chalmers University of Technology, Onsala Space Observatory, SE-439 92 Onsala, Sweden}

\author{S. B.~Cenko}
\affiliation{Astrophysics Science Division, NASA Goddard Space Flight Center, Mail Code 661, Greenbelt, Maryland 20771, USA}
\affiliation{Joint Space-Science Institute, University of Maryland, College Park, MD 20742, USA}

\author{D.~Frail}
\affiliation{National Radio Astronomy Observatory, P.O.\,Box O, Socorro, NM 87801, USA}

\author{S. R.~Kulkarni}
\affiliation{California Institute of Technology, 1200 E California Blvd. MC 249-17, Pasadena, CA 91125, USA}

\author{S.~Myers}
\affiliation{National Radio Astronomy Observatory, P.O.\,Box O, Socorro, NM 87801, USA}

\correspondingauthor{Marin M. Anderson}
\email{mmanders@astro.caltech.edu}

\begin{abstract}
We present the discovery of a nuclear transient with the Caltech--NRAO Stripe\,82 Survey (CNSS), a dedicated radio transient survey carried out with the Karl G.~Jansky Very Large Array (VLA). This transient, CNSS J001947.3+003527, exhibited a turn-on over a timescale of $\lesssim$1\,yr, increasing in flux density at 3\,GHz from $<0.14\,\rm{mJy}$ in 2014 February to $4.4\pm0.1\,\rm{mJy}$ in 2015 March, reaching a peak luminosity of $5\times10^{28}\,\text{erg\,s}^{-1}\,\text{Hz}^{-1}$ around 2015 October. The association of CNSS J0019+00 with the nucleus (Gaia and our very long baseline interferometry positions are consistent to within 1\,pc) of a nearby S0 Seyfert galaxy at 77 Mpc, together with the radio spectral evolution, implies that this transient is most likely a tidal disruption event (TDE). Our equipartition analysis indicates the presence of a $\sim$15,000\,km\,s$^{-1}$ outflow, having energy $\sim$10$^{49}$ erg. We derive the radial density profile for the circumnuclear material in the host galaxy to be proportional to $R^{-2.5}$. All of these properties suggest resemblance with radio-detected thermal TDEs like ASASSN-14li and XMMSL1 J0740-85. No significant X-ray or optical emission is detected from CNSS J0019+00, although this may simply be due to the thermal emission being weak during our late-time follow-up observations. From the CNSS survey we have obtained the first unbiased measurement of the rate of radio TDEs, $R(>500\,\mu{\rm Jy})$ of about $2\times10^{-3}$\,deg$^{-2}$, or equivalently a volumetric rate of about 10\,Gpc$^{-3}$\,yr$^{-1}$. This rate implies that all-sky radio surveys such as the VLA Sky Survey and those planned with ASKAP, will find many tens of radio TDEs over the next few years.
\end{abstract}

\section{Introduction}\label{introduction}
The passage of a star within the tidal radius of a galaxy's central supermassive black hole (SMBH) results in the disruption of the star by the tidal field of the SMBH. The distance from the SMBH at which this tidal disruption event (TDE) occurs is a function of the mass of the SMBH, as well as the mass and radius of the disrupted star~\citep{Hills1975}. For a (non-spinning) SMBH with mass $M \lesssim 10^8\,M_\odot$ disrupting a solar type star, this distance is greater than the Schwarzchild radius of the BH and the stellar material will not be swallowed whole. Instead, approximately half of the stellar mass is accreted onto the SMBH, while the other half is ejected, and a luminous flare is produced that can be observed across the electromagnetic spectrum.

Theoretical predictions for TDEs initially recognized the potential of these events for verifying the existence of SMBHs. Later they highlighted TDEs as probes of the central regions of galaxies: around SMBHs that were usually otherwise dormant~\citep{Hills1975, Rees1988, Phinney1989}. TDEs were predicted to result in bright flares with super-Eddington luminosities peaking in the soft X-ray and UV bands. These theoretical predictions were found to be consistent with the observational behavior of thermal TDEs, which were first discovered in the soft X-rays, with subsequent discoveries in the soft X-rays and UV~(see~\citealt{Komossa2015} and~\citealt{Lodato+2015} for reviews of TDE observations and theory, respectively). The discovery of \textit{Swift} J1644+57~\citep{Levan+2011}, with its extremely luminous hard X-ray~\citep{Burrows+2011} and radio~\citep{Zauderer+2011, Berger+2012} emission indicative of a relativistic outflow from a collimated jet, revealed the existence of non-thermal TDEs~\citep{Bloom+2011}, and their ability to serve as a laboratory for testing the connection between accretion and launching of jets, as well as for measuring SMBH spin~\citep{GIannios+Metzger2011, vanVelzen+2011}.

Since the first discovery of TDEs with ROSAT~\citep{Bade+1996}, there are now more than 90 confirmed or candidate TDEs\footnote{https://tde.space}, with discoveries being made increasingly in optical surveys, as well as UV and X-rays. Of the known TDEs, seven have definitive radio detections, including \textit{Swift} J1644+57~\citep{Zauderer+2011, Berger+2012, Zauderer+2013, Eftekhari+2018} and two additional jetted, non-thermal TDEs, \textit{Swift} J2058+05~\citep{Cenko+2012,Brown+2017b} and \textit{Swift} J1112.2-8238~\citep{Brown+2015,Brown+2017b}. IGR J12580+0134 is also proposed to be a non-thermal TDE, with a relativistic jet that is viewed off-axis~\citep{Nikolajuk+Walter2013, Irwin+2015, Lei+2016}. The TDE Arp 299-B\,AT1 was initially detected as a near-infrared transient; radio very-long baseline interferometry (VLBI) resolved a relativistic TDE-driven jet~\citep{mattila2018}. More recently, the thermal TDEs ASASSN-14li~\citep{Alexander+2016,Krolik+2016,vanVelzen+2016,Yalinewich+2019} and XMMSL1 J0750-85~\citep{Alexander+2017} were found to have radio emission that indicated sub-/non-relativistic outflows, indicating that these are the first thermal TDEs with detected radio emission. These objects revealed the possibility that all TDEs are accompanied by radio emission, with earlier non-detections~\citep{Bower2011, Bower+2013, vanVelzen+2013, Arcavi+2014, Chornock+2014} a result of their lower radio luminosities: both ASASSN-14li and XMMSL1 J0750-85 were relatively nearby at $z\lesssim0.02$, whereas the median TDE redshift is approximately $z\sim0.1$~\citep{Komossa2015} --- though see \citealt{Blagorodnova+2017} and \citealt{Saxton+2019} for upper limits on the radio luminosities of TDEs that are more than an order of magnitude below that of ASASSN-14li in hosts that are similarly nearby. While the number of detected thermal TDEs is expected to grow rapidly given the rise of time-domain optical surveys (e.g., ZTF, ASAS-SN, Pan-STARRS, and eventually LSST), the recent radio detection of TDEs indicates the excellent potential for radio surveys to contribute to the TDE discovery space. With the radio TDE detection rate potentially unaffected by obscuration, such TDEs can therefore offer a complementary view of TDE event rates and host galaxies.

Here we present the discovery of the radio transient CNSS J001947.3+003527 (hereafter CNSS J0019+00) identified in the Caltech--NRAO Stripe\,82 Survey (CNSS), and located in the nucleus of a nearby ($z=0.018$) galaxy. Its association with the nucleus suggests that CNSS J0019+00 is likely a TDE, the first radio-discovered event of its kind. The rest of this paper is organized as follows. In Section~\ref{observations} we describe the discovery of CNSS J0019+00 and the subsequent follow-up observations at radio, X-ray, and optical wavelengths. Section~\ref{analysis} describes the modeling of the follow-up radio spectra, and how they point to a Newtonian expanding outflow. Section~\ref{sec:host} describes the host galaxy of CNSS J0019+00. We conclude with a summary and discussion in Section~\ref{discussion}.

\section{Observations}\label{observations}

\subsection{Caltech--NRAO Stripe\,82 Survey}\label{cnss}
CNSS is a five-epoch survey with the NSF's Karl G.~Jansky Very Large Array~\citep[VLA;][]{Perley+2011} at \textit{S} band (2--4\,GHz) that was carried out between 2013 December and 2015 May. It was designed to probe timescales of days, months, and years, and thus significantly advance the understanding of slow transient phenomena in the radio sky. Each epoch of CNSS covers the full $\sim$270\,deg$^{2}$ Sloan Digital Sky Survey (SDSS) Stripe\,82 region with a uniform single-epoch sensitivity of $\sim$80\,$\mu$Jy and a spatial resolution of $\sim$3$''$. Through the use of On The Fly (OTF) mosaicing~\citep{Mooley+2018,Mooley+2019}, the CNSS combines shallow mapping of the sky with the excellent sensitivity of the VLA, thus delivering a high survey speed without being compromised by large slew-and-settle overheads. The results from the 50\,deg$^{2}$ pilot survey were presented in~\cite{Mooley+2016}, and included the discovery of radio transients consistent with an RS CVn binary and a dKe star. CNSS was designed as a pathfinder for wide-field radio surveys, including the VLA Sky Survey~\citep[VLASS;][]{lacy2019}.

\subsection{CNSS J001947.3+003527}\label{tde3}
CNSS J0019+00 was first detected in CNSS epoch 4 on 2015 March 21 at a 3\,GHz flux density of $4.4 \pm 0.1$\,mJy at the position (R.A., decl.)~=~($00^\text{h}19^\text{m}47^\text{s}.3, +00^\circ35'27''$). It was not detected in the first three epochs of CNSS between 2013 December and 2014 February at a $3\sigma$ upper limit of 0.14\,mJy in the combined epochs 1--3 coadded image, and was therefore identified as a candidate transient event. There is no coincident source in the Faint Images of the Radio Sky at Twenty-cm~\citep[FIRST;][]{Becker+1995} survey, with a 0.5\,mJy $3\sigma$ upper limit at 1.4\,GHz (mean epoch 1999) at the location of CNSS J0019+00. By CNSS epoch 5 on 2015 April 19, the transient had increased to a 3\,GHz flux density of $5.1\pm0.1$\,mJy.

\begin{figure}[htb!]
\begin{center}
	\includegraphics[width=0.6\textwidth]{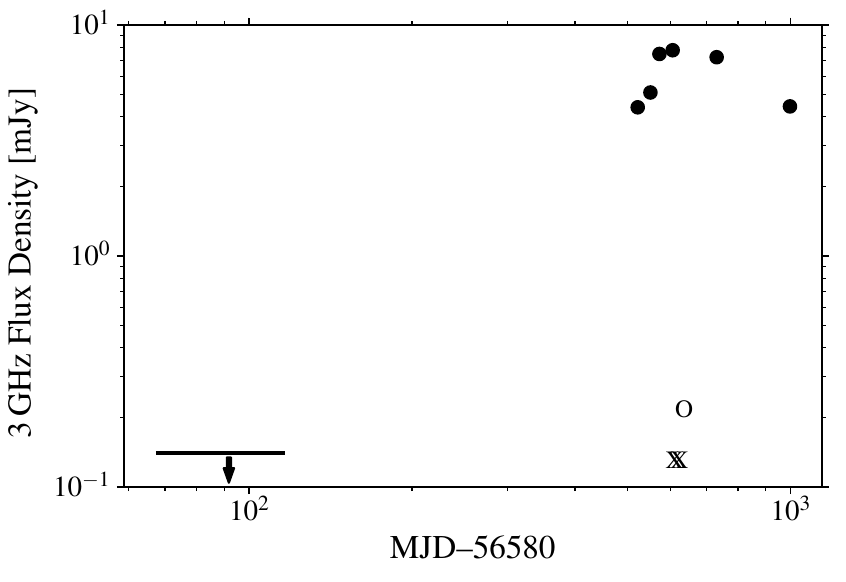}
	\caption{CNSS J0019+00 3\,GHz light curve. The 0.14\,mJy upper limit is from the non-detection in CNSS epochs 1--3. The 3 GHz flux densities are from CNSS epochs 4 and 5, and 4 follow-up observations of CNSS J0019+00 with the VLA spanning approximately 1.5\,yr post-discovery. The Xs and O mark the dates of follow-up \textit{Swift} and Keck II DEIMOS observations, respectively. The dates are referenced to the approximate explosion date on MJD 56580, as determined by fitting the radio SEDs (see Section~\ref{analysis}).}\label{fig:lightcurve}
\end{center}
\end{figure}

One method by which extragalactic transients were identified in CNSS was through the cross-matching of radio transient candidates with galaxy catalogs for the local universe, which are typically developed for gravitational-wave event follow-up. For this work, the Census of the Local Universe~\citep[CLU;][]{Cook+2019} was used to identify objects out to a volume of 200\,Mpc, for which we expect a very low false-positive rate due to background active galactic nuclei (AGNs). Candidate transients identified in CNSS were cross-matched with CLU galaxies using a $30$\,kpc projected radius for each galaxy, with the aim of identifying radio transients associated with explosive events, including TDEs, supernovae (SNe), and gamma-ray burst (GRB) orphan afterglows. CNSS J0019+00 was the only significant extragalactic transient identified on timescales shorter than the duration of the survey, rather than through comparison to existing surveys (e.g., FIRST). Cross-matching with CLU and SDSS established that CNSS J0019+00 is coincident with the nucleus of a Seyfert 2 galaxy (SDSS J0019+00) at a distance of 77.1\,Mpc ($z=0.018$). After its discovery, follow-up observations of CNSS J0019+00 were carried out at radio, X-ray, and optical wavelengths. Figure~\ref{fig:lightcurve} shows the 3\,GHz light curve and observation timeline for CNSS J0019+00.

\subsection{VLA Observations}\label{radioobs}
Following the discovery of CNSS J0019+00 in 2015 March, we continued to monitor the source with the VLA over the course of the next 14 months (under program codes 15A-421, PI: Gregg Hallinan; 15B-364, PI: Kunal Mooley; and 16A-237, PI: Shri Kulkarni). It was observed from \textit{L} through \textit{Ku}~band (1--16\,GHz) in order to fully sample the spectrum of the source across four follow-up epochs, spanning 2015 May 10 -- 2016 July 8. The follow-up spectra of CNSS J0019+00 are well fit by a slowly evolving synchrotron spectrum, with both the peak frequency and peak flux of the spectrum declining over time. 

\begin{figure*}[htb!]
\centering
	\includegraphics[width=1\textwidth]{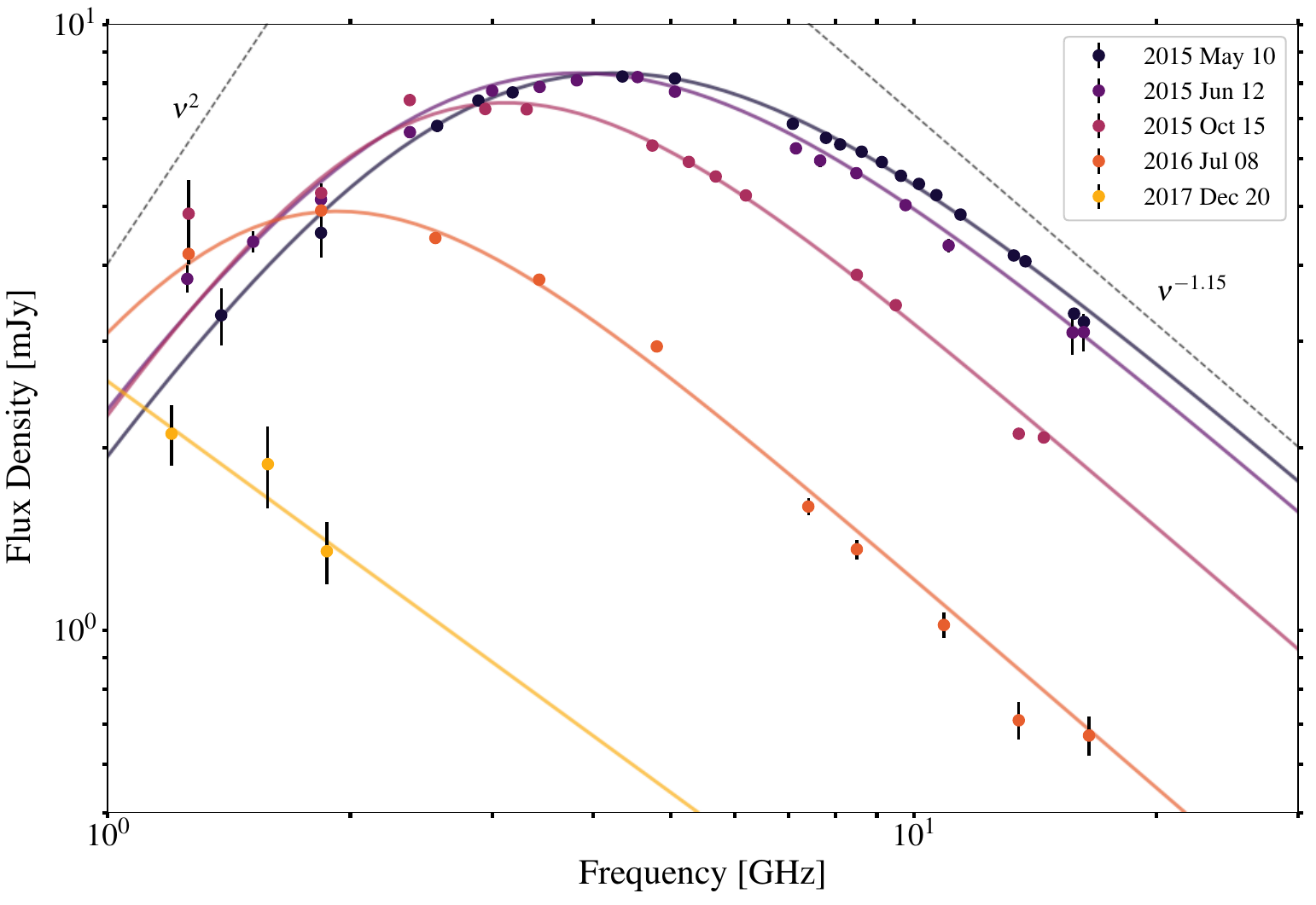}
	\caption{Evolution of the broadband spectral energy distribution (SED) of CNSS J0019+00, as observed with the VLA in five epochs spaced across 2.5\,yr. The synchrotron spectra are modeled according to Equation 1 of \citealt{Granot+Sari2002}. In our analysis, we assume optically thick and thin spectral indices of $\nu^2$ and $\nu^{-1.15}$, respectively. The latter is expected for an electron energy distribution described by a power law with $p=3.3$.}
	\label{fig:vlafollowup}
\end{figure*}

Table~\ref{tab:radioobs} summarizes the radio follow-up measurements, which are plotted in Figure~\ref{fig:vlafollowup}. These also include measurements of the flux of CNSS J0019+00 at \textit{L}~band (1--2\,GHz) from observations on 2017 December 20 (program code 17B-409, PI: Gregg Hallinan). For all observations, 3C48 served as the absolute flux and bandpass calibrator. Phase calibration was done using one of J0022+0014, J006-0623, J0022+0608, J0015-0127. All data reduction was done in the Common Astronomy Software Applications~\citep[CASA;][]{McMullin+2007}.

\begin{small}
\begin{longtable}{|r|C|C|}
	\caption{VLA observations of CNSS J0019+00.}\label{tab:radioobs} \\

	\hline \multicolumn{1}{|c|}{\textbf{UT Date}}	& \multicolumn{1}{c|}{\textbf{Frequency (GHz)}}		& \multicolumn{1}{c|}{\textbf{Flux Density (mJy)}} \\ \hline

	\endfirsthead
	
	\multicolumn{3}{c}%
	{{\bfseries \tablename\ \thetable{} -- continued from previous page}} \\
	\hline \multicolumn{1}{|c|}{\textbf{UT Date}} 	& \multicolumn{1}{c|}{\textbf{Frequency (GHz)}} 	& \multicolumn{1}{c|}{\textbf{Flux Density (mJy)}} \\ \hline 
\endhead

\hline \multicolumn{3}{|r|}{{Continued on next page}} \\ \hline
\endfoot

\hline \hline
\endlastfoot
    \multicolumn{1}{|l|}{(E1) 2013 Dec 29} & 2.9 \pm 0.9 & <0.296 \\
    \multicolumn{1}{|l|}{(E2) 2014 Jan 11} & 2.9 \pm 0.9 & <0.279 \\
    \multicolumn{1}{|l|}{(E3) 2014 Feb 17} & 2.9 \pm 0.9 & <0.276 \\
    \multicolumn{1}{|l|}{(E4) 2015 Mar 21} & 2.9 \pm 0.9 & 4.4 \pm 0.1 \\
    \multicolumn{1}{|l|}{(E5) 2015 Apr 19} & 2.9 \pm 0.9 & 5.1 \pm 0.1 \\
    \hline
	2015 May 10.53 & 1.4 \pm 0.15 & 3.31 \pm 0.36 \\
	May 10.53 & 1.8 \pm 0.15 & 4.53 \pm 0.41 \\
	May 10.53 & 2.6 \pm 0.15 & 6.80 \pm 0.16 \\
	May 10.53 & 2.9 \pm 0.15 & 7.49 \pm 0.11 \\
	May 10.53 & 3.2 \pm 0.15 & 7.72 \pm 0.07 \\
	May 10.55 & 4.4 \pm 0.40 & 8.21 \pm 0.08 \\
	May 10.55 & 5.1 \pm 0.40 & 8.14 \pm 0.04 \\
	May 10.55 & 7.1 \pm 0.40 & 6.85 \pm 0.04 \\
	May 10.55 & 7.8 \pm 0.40 & 6.50 \pm 0.05 \\
	May 10.55 & 8.1 \pm 0.25 & 6.34 \pm 0.04 \\
	May 10.55 & 8.6 \pm 0.25 & 6.17 \pm 0.04 \\
	May 10.55 & 9.1 \pm 0.25 & 5.93 \pm 0.04 \\
	May 10.55 & 9.6 \pm 0.25 & 5.63 \pm 0.04 \\
	May 10.55 & 10.2 \pm 0.25 & 5.45 \pm 0.04 \\
	May 10.55 & 10.7 \pm 0.25 & 5.23 \pm 0.04 \\
	May 10.55 & 11.4 \pm 0.25 & 4.85 \pm 0.04 \\
	May 10.57 & 13.3 \pm 0.25 & 4.16 \pm 0.05 \\
	May 10.57 & 13.8 \pm 0.25 & 4.06 \pm 0.04 \\
	May 10.57 & 15.8 \pm 0.25 & 3.33 \pm 0.05 \\
	May 10.57 & 16.3 \pm 0.25 & 3.23 \pm 0.04 \\
	\hline
	2015 Jun 12.71 & 1.3 \pm 0.15 & 3.80 \pm 0.20 \\
	Jun 12.71 & 1.5 \pm 0.15 & 4.38 \pm 0.17 \\
	Jun 12.71 & 1.8 \pm 0.15 & 5.14 \pm 0.17 \\
	Jun 12.72 & 2.4 \pm 0.15 & 6.64 \pm 0.08 \\
	Jun 12.72 & 3.0 \pm 0.15 & 7.77 \pm 0.06 \\
	Jun 12.72 & 3.4 \pm 0.15 & 7.88 \pm 0.06 \\
	Jun 12.72 & 3.8 \pm 0.15 & 8.08 \pm 0.07 \\
	Jun 12.72 & 4.5 \pm 0.40 & 8.19 \pm 0.10 \\
	Jun 12.72 & 5.1 \pm 0.40 & 7.74 \pm 0.10 \\
	Jun 12.72 & 7.1 \pm 0.40 & 6.24 \pm 0.14 \\
	Jun 12.72 & 7.7 \pm 0.40 & 5.96 \pm 0.15 \\
	Jun 12.73 & 8.5 \pm 0.25 & 5.68 \pm 0.08 \\
	Jun 12.73 & 9.8 \pm 0.25 & 5.03 \pm 0.10 \\
	Jun 12.73 & 11.0 \pm 0.25 & 4.32 \pm 0.11 \\
	Jun 12.74 & 15.7 \pm 0.25 & 3.10 \pm 0.25 \\
	Jun 12.74 & 16.3 \pm 0.25 & 3.10 \pm 0.22 \\
	\hline
	2015 Oct 15.32 & 1.3 \pm 0.15 & 4.87 \pm 0.66 \\
	Oct 15.32 & 1.8 \pm 0.25 & 5.27 \pm 0.20 \\
	Oct 15.32 & 2.4 \pm 0.25 & 7.50 \pm 0.17 \\
	Oct 15.32 & 2.9 \pm 0.25 & 7.25 \pm 0.10 \\
	Oct 15.32 & 3.3 \pm 0.25 & 7.24 \pm 0.09 \\
	Oct 15.33 & 4.7 \pm 0.40 & 6.31 \pm 0.06 \\
	Oct 15.33 & 5.3 \pm 0.40 & 5.93 \pm 0.05 \\
	Oct 15.33 & 5.7 \pm 0.40 & 5.61 \pm 0.05 \\
	Oct 15.33 & 6.2 \pm 0.40 & 5.22 \pm 0.06 \\
	Oct 15.33 & 8.5 \pm 0.25 & 3.86 \pm 0.04 \\
	Oct 15.33 & 9.5 \pm 0.25 & 3.44 \pm 0.04 \\
	Oct 15.34 & 13.5 \pm 0.25 & 2.11 \pm 0.04 \\
	Oct 15.34 & 14.5 \pm 0.25 & 2.08 \pm 0.03 \\
	
	\hline
	2016 Jul 08.60 & 1.3 \pm 0.15 & 4.18 \pm 0.16 \\
	Jul 08.60 & 1.8 \pm 0.15 & 4.93 \pm 0.14 \\
	Jul 08.60 & 2.5 \pm 0.15 & 4.44 \pm 0.07 \\
	Jul 08.60 & 3.4 \pm 0.15 & 3.79 \pm 0.06 \\
	Jul 08.61 & 4.8 \pm 0.40 & 2.94 \pm 0.05 \\
	Jul 08.61 & 7.4 \pm 0.40 & 1.60 \pm 0.05 \\
	Jul 08.61 & 8.5 \pm 0.25 & 1.36 \pm 0.05 \\
	Jul 08.61 & 10.9 \pm 0.25 & 1.02 \pm 0.05 \\
	Jul 08.61 & 13.5 \pm 0.25 & 0.71 \pm 0.05 \\
	Jul 08.61 & 16.5 \pm 0.25 & 0.67 \pm 0.05 \\
	\hline
	2017 Dec 20.03 & 1.20 \pm 0.15 & 2.11 \pm 0.24 \\ 
	Dec 20.03 & 1.58 \pm 0.15 & 1.88 \pm 0.29 \\
	Dec 20.03 & 1.87 \pm 0.15 & 1.35 \pm 0.16 \\
\end{longtable}
\tablecomments{CNSS J0019+00 $3\sigma$ flux density upper limits from CNSS epochs 1--3 and measurements from epochs 4 and 5, as well as VLA follow-up observations with SEDs spanning approximately 1--16\,GHz. The $3\sigma$ upper limit at the location of CNSS J0019+00 from the combined epochs 1--3 coadd is 0.14\,mJy. The follow-up observations listed here span approximately 573 to 1527\,d post-outburst, with the time since outburst determined from model fitting of the individual SEDs (see Section~\ref{analysis}). Uncertainties on the peak flux density measurements are from the image RMS noise.}
\end{small}

\subsection{VLBA Observations}\label{radiovlbiobs}
We conducted Very Long Baseline Array (VLBA) observations at the location of CNSS J0019+00 on 2015 November 10 and 2016 August 31, in order to place a constraint on, or potentially resolve, a relativistically expanding jet (in the event of a non-thermal TDE scenario, e.g. \textit{Swift} J1644+57). Observations were conducted at 4.38 and 7.40\,GHz, as part of the Director's Discretionary Time (project code BM444), with J02253+1608 as the fringe finder source and J0016-0015 for phase calibration. Due to the limited Local Sidereal Time (LST) range accessible for observing equatorial sources with the VLBA, the observations were split across the 2 epochs, with observation blocks of 2.5 hr each (see Table~\ref{tab:vlbaobs}).

The source is unresolved in both VLBA epochs, consistent with the source size $\lesssim 1\times10^{17}$\,cm as determined by modeling of the radio spectral energy distribution (SED; see Section~\ref{analysis}). From \textit{Gaia}, the nucleus of the host galaxy is (R.A., decl.)~=~($00^\text{h}19^\text{m}47^\text{s}.33493, +00^\circ35'26''.8126$) $\pm$ (0.34, 0.26) milliarcsec~\citep{GaiaCollaboration+2016, GaiaCollaboration+2018, Lindegren+2018}. From the VLBA observations, the best-fit position of CNSS J0019+00 is (R.A., decl.)~=~($00^\text{h}19^\text{m}47^\text{s}.335038, +00^\circ35'26''.8102$) $\pm$ (0.12, 0.04) milliarcsec. This is coincident with the \textit{Gaia} position to within $\sim$3\,mas, implying that CNSS J0019+00 is consistent with the nuclear region of the host galaxy to within $\sim$1\,pc.

\begin{small}
\begin{longtable}{|r|C|C|}
	\caption{VLBA observations of CNSS J0019+00.}\label{tab:vlbaobs} \\

	\hline \multicolumn{1}{|c|}{\textbf{UT Date}}	& \multicolumn{1}{c|}{\textbf{Frequency (GHz)}}		& \multicolumn{1}{c|}{\textbf{Flux Density (mJy)}} \\ \hline

	\endfirsthead
	
	\multicolumn{3}{c}
	{{\bfseries \tablename\ \thetable{} -- continued from previous page}} \\
	\hline \multicolumn{1}{|c|}{\textbf{UT Date}} 	& \multicolumn{1}{c|}{\textbf{Frequency (GHz)}} 	& \multicolumn{1}{c|}{\textbf{Flux Density (mJy)}} \\ \hline 
\endhead

\hline \multicolumn{3}{|r|}{{Continued on next page}} \\ \hline
\endfoot

\hline \hline
\endlastfoot
	2015 Nov 10.15 & 4.37 \pm 0.06 & 4.67 \pm 0.06 \\
	Nov 10.15 & 7.40 \pm 0.06 & 2.95 \pm 0.08 \\
    \hline
	2016 Aug 31.35 & 4.37 \pm 0.06 & 2.33 \pm 0.07 \\
	Aug 31.35 & 7.40 \pm 0.06 & 1.07 \pm 0.07 \\    
\end{longtable}
\end{small}

\subsection{\textit{Swift} Observations}\label{x-rayobs}
Neil Gehrels Swift Observatory~\citep{Gehrels+2004, Burrows+2005} X-ray observations were triggered within approximately 2 weeks of the first radio follow-up observation (see Table~\ref{tab:swiftobs}). No X-ray counterpart was detected in either exposure, and the combined upper limit (90\% confidence) to the count rate in the 0.3--10\,keV soft X-ray band is $9.3\times10^{-4}\,\text{count\,s}^{-1}$. This corresponds to an upper limit in the X-ray luminosity of less than $2.4\times10^{40}\,\text{erg\,s}^{-1}$ assuming a power law with photon index 2 (typical for non-thermal emission), or an upper limit of less than $4.0\times10^{40}\,\text{erg\,s}^{-1}$ assuming blackbody emission that peaks at 10,000\,K (or 1.2 keV, typical of a disk formed after a TDE). Additionally, the \textit{U} band magnitudes from \textit{Swift} UVOT~\citep{Roming+2005}, $17.39\pm0.02$ and $17.36\pm0.02$ (AB magnitude), did not change significantly between the two epochs.

\begin{deluxetable}{cCCCCC}[htb!]
	\tabletypesize{\scriptsize}
	\tablecolumns{6}
	\tablecaption{\textit{Swift} observations of CNSS J0019+00.\label{tab:swiftobs}}
	\tablehead{
					&						&					&								& \colhead{Luminosity ($10^{40}$ erg s$^{-1}$)} \vspace{-0.4cm} \hspace{-4.5cm}		\\
	\colhead{UT Date}	& \colhead{Exposure Time (ks)}	& \colhead{Band (keV)}	& \colhead{Count Rate (counts s$^{-1}$)}	& \vspace{-0.4cm}												&\\
					&						&					&								& \colhead{Power Law with Photon Index 2} & \colhead{10,000 K Blackbody}
	}
	\startdata
	\vspace{-0.2cm} 2015 May 26 	& 5.8 &  						&  		&				\\
							&	 & \vspace{-0.2cm}  0.3-10	& \leq 9.3 \times 10^{-4} 	& \leq 2.4  & \leq 4.0 \\
	2015 Jun 07  				& 6.3 &						&		&		  		\\
	\enddata
	\tablecomments{\textit{Swift} follow-up observations at the location of CNSS J0019+00, starting approximately 464 days after the outflow launch.}
\end{deluxetable}

\subsection{Keck II DEIMOS Observations}\label{opticalobs}
\begin{figure}[htb!]
\centering
	\includegraphics[width=1\textwidth]{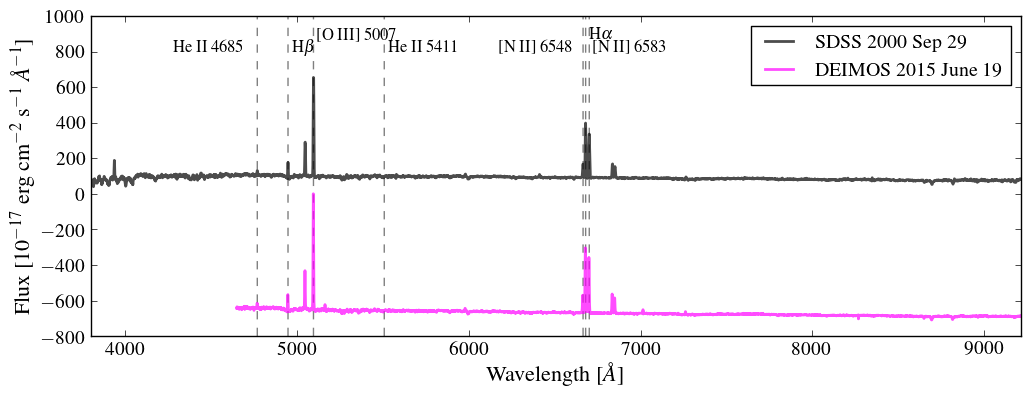}
	\caption{Optical spectra of the host galaxy of CNSS J0019+00, with SDSS in 2000 September 29 (black line) and DEIMOS on Keck II in 2015 June 19 (magenta line) approximately 612 days post-outburst. The host is a Seyfert 2 galaxy, based on the measured nebular line flux ratios and high surface brightness nucleus. See \S\ref{sec:host} for details.}
	\label{fig:opticalspec}
\end{figure}

Optical observations of the host galaxy were conducted on 2015 June 19, approximately 1.5 months after the first radio follow-up observation, with the DEep Imaging Multi-Object Spectrograph~\citep[DEIMOS;][]{Faber+2003} on Keck II. The spectrum has a resolution of 1.3\,\AA, and was reduced using standard DEIMOS tasks in IRAF and IDL, including wavelength calibration using arc lamps and flux calibration using standard stars. Figure~\ref{fig:opticalspec} shows our follow-up spectrum of the host galaxy as compared with an SDSS spectrum of the host from 2000 September 29. No significant change in the spectrum is observed before and after the transient event, indicating that either there is no associated optical transient or that any optical signatures had faded by the time of our follow-up observations approximately 612 days after the launch of the outflow (see Section~\ref{analysis}).

\section{Modeling of the Synchrotron Spectra}\label{analysis}
The spectra of CNSS J0019+00 (Figure~\ref{fig:vlafollowup}) are well described by synchrotron emission from an outflow expanding into and shocking the surrounding medium. From the evolution of the synchrotron spectra observed from CNSS J0019+00 in the radio follow-up observations, a number of parameters characterizing the source can be derived as a function of time, including the size of the source, the minimal equipartition energy, the ambient density, and magnetic field strength. Each spectrum provides an independent constraint on these parameters, based only on the frequency and flux at the peak of the spectrum and the electron power-law index ($p$, where $N_e(\gamma) \propto \gamma^{-p}$ for $\gamma \geq \gamma_m$, and $\gamma_m$ is the Lorentz factor of the lowest energy electrons in the distribution), which determines the slope of the optically thin half of the spectrum. This procedure has been well established and used to study both relativistically and non-relativistically expanding outflows, including for GRBs~\citep[e.g.,][]{Wijers+Galama1999}, SNe~\citep[ e.g.,][]{Chevalier1998}, and TDEs~\citep[e.g.,][]{Zauderer+2011}.

\begin{deluxetable}{CCCCCCCCC}[htb!]
	\tablecolumns{9}
	\tablecaption{CNSS J0019+00 micro and macrophysical parameters\label{tab:bestfitparams}}
	\tablehead{
	(1) & (2) & (3) & (4) & (5) & (6) & (7) & (8) & (9) \\
	\colhead{$\Delta t$} & \colhead{$\nu_p$} & \colhead{$F_{\nu_p}$} & \colhead{$s$} & \colhead{$R_{\text{eq}}$} & \colhead{$E_{\text{eq}}$} & \colhead{$n_e$} & \colhead{$B$} & \colhead{$M_\text{ej}$} \\
	\colhead{(d)} & \colhead{(GHz)} & \colhead{(mJy)} & & \colhead{($10^{16}$\,cm)} & \colhead{($10^{49}$\,erg)} & \colhead{($10^3$\,cm$^{-3}$)} & \colhead{(G)} & \colhead{($10^{-3}\,M_\odot$)}
	}
	\startdata
	 573 	 & 4.26 \pm 0.04 	 & 8.27 \pm 0.03 	 & 0.64 \pm 0.02 	 & 7.23 \pm 0.06 	 & 1.37 \pm 0.01 	 &  3.98 \pm 0.07 	 & 0.380 \pm 0.003 	 & 6.42 \pm 0.06 \\
	 606 	 & 3.87 \pm 0.03 	 & 8.08 \pm 0.04 	 & 0.64 \pm 0.02 	 & 7.87 \pm 0.07 	 & 1.46 \pm 0.02 	 &  3.31 \pm 0.06 	 & 0.347 \pm 0.003 	 & 6.86 \pm 0.07 \\
	 731 	 & 3.12 \pm 0.04 	 & 7.41 \pm 0.06 	 & 0.86 \pm 0.04 	 & 9.38 \pm 0.14 	 & 1.64 \pm 0.03 	 &  2.18 \pm 0.06 	 & 0.281 \pm 0.004 	 & 7.7 \pm 0.1 \\
	 998 	 & 1.92 \pm 0.06 	 & 4.95 \pm 0.09 	 & 0.91 \pm 0.08 	 & 12.59 \pm 0.38 	 & 1.63 \pm 0.06 	 &  0.90 \pm 0.05 	 & 0.180 \pm 0.005 	 & 7.6 \pm 0.3 \\
	 \enddata
	\tablecomments{Table of micro and macrophysical parameters for CNSS J0019+00, based on model fitting of the radio SEDs. The values were computed with $p=3.3$, and under the assumption of $\epsilon_e = \epsilon_b = 1/3$, and assuming a filling factor of $f=0.5$ (see Appendix~\ref{ap1}). At each epoch, the expansion velocity is consistent with $\beta_\text{ej} \approx 0.05$. The ejecta mass $M_\text{ej}$ is computed by approximating the equipartition energy with the kinetic energy of the outflow; because the outflow has not yet decelerated at these epochs, the equipartition energy is a lower limit on the kinetic energy, and the ejecta mass should be interpreted as a lower limit.}
\end{deluxetable}

\begin{figure}[htb!]
\centering
	\subfigure[]{\includegraphics[width=0.45\textwidth]{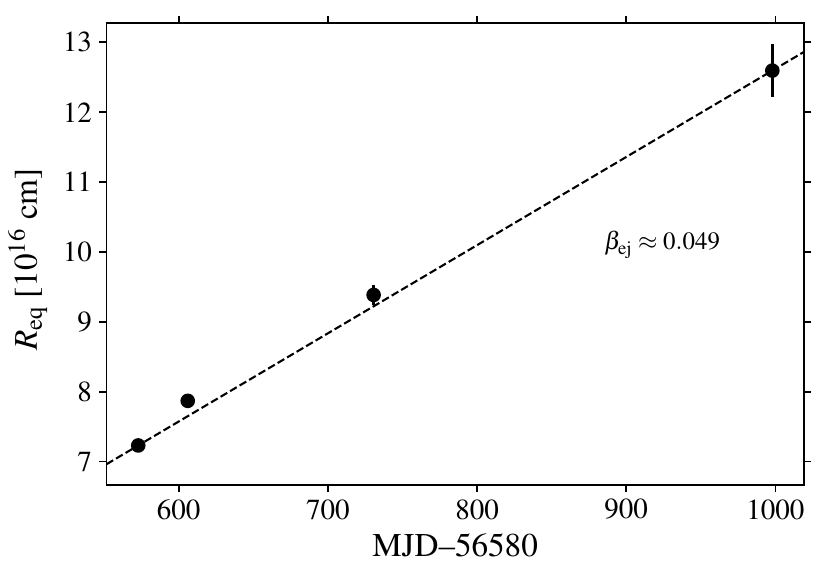}}
	\subfigure[]{\includegraphics[width=0.45\textwidth]{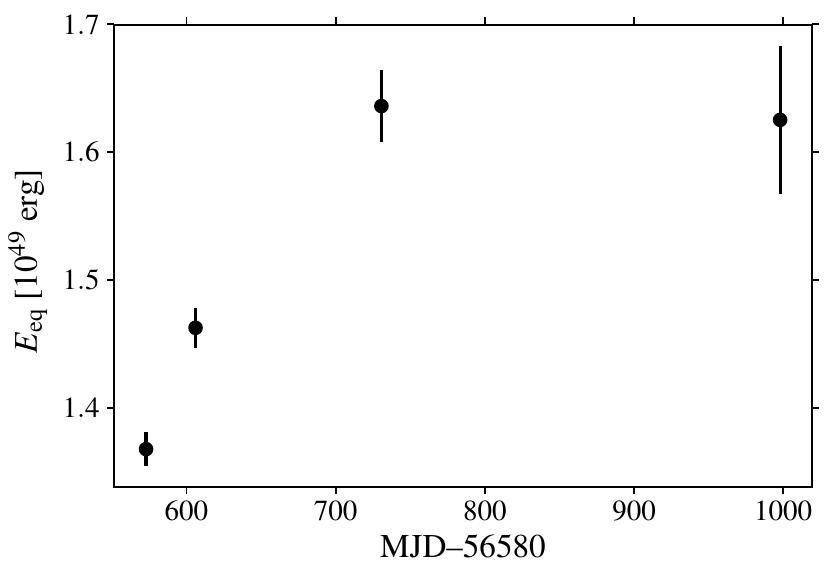}}
	\subfigure[\label{fig:density}]{\includegraphics[width=0.65\textwidth]{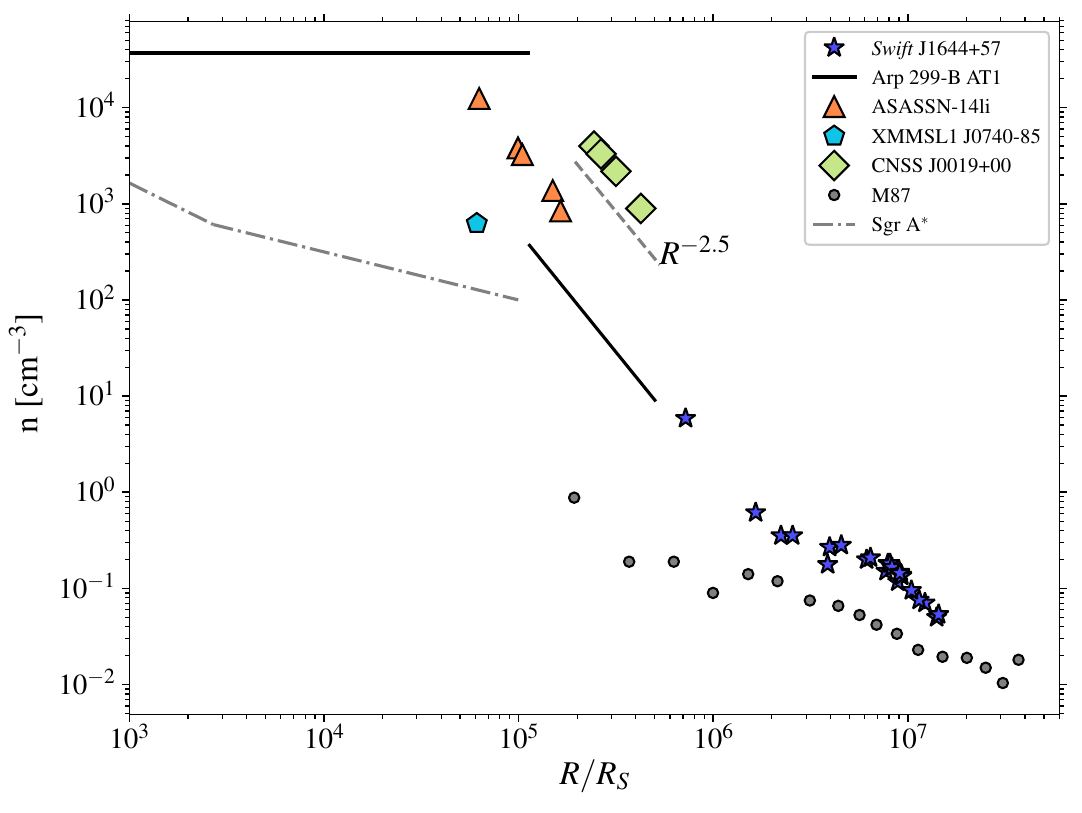}}
	\caption{Equipartition radius (a) and energy (b) as a function of time since outflow, as derived from the radio SEDs (see Appendix~\ref{ap1}). Each pair of ($R_\text{eq}$, $E_\text{eq}$) values are independent of the others, calculated using only $F_p$, $\nu_p$, and $p$ as derived from each follow-up observation. Jointly, the equipartition radii from each epoch can therefore provide a strong constraint on the outflow velocity, which for CNSS J0019+00 is constant, at $\sim15,000\,\text{km\,s}^{-1}$. We also derive the ambient density as a function of the equipartition radius, scaled to the Schwarzchild radius of the SMBH, and compare with other radio-detected TDEs as well as M87 and Sgr A$^*$ (c). A SMBH mass of $\sim10^6\,M_\odot$ is used for CNSS J0019+00, \textit{Swift} J1644+57, and ASASSN-14li. For XMMSL1 J0740-85, we use a SMBH mass of $3.5\times10^6\,M_\odot$~\citep{Saxton+2017}. The density and radius values for the thermal TDEs, ASASSN-14li and XMMSL1 J0740-85, have been recomputed according to our method and set of assumptions as outlined in Appendix~\ref{ap1}. The dashed line shows the circumnuclear density profile of $R^{-2.5}$ inferred from our observations. Data are from -- for \textit{Swift} J1644+57, \citealt{Eftekhari+2018}; for the circumnuclear density derived from modeling the radio jet of Arp 299B-AT1, \citealt{mattila2018}; for ASASSN-14li, \citealt{Alexander+2016}; for XMMSL1 J0740-85, \citealt{Alexander+2017}; for M87, \citealt{russell2018}; for Sgr A$^*$, \citealt{gillessen2019}.}
	\label{fig:equipartition}
\end{figure}

\begin{figure}[htb!]
\centering
	\includegraphics[width=1\textwidth]{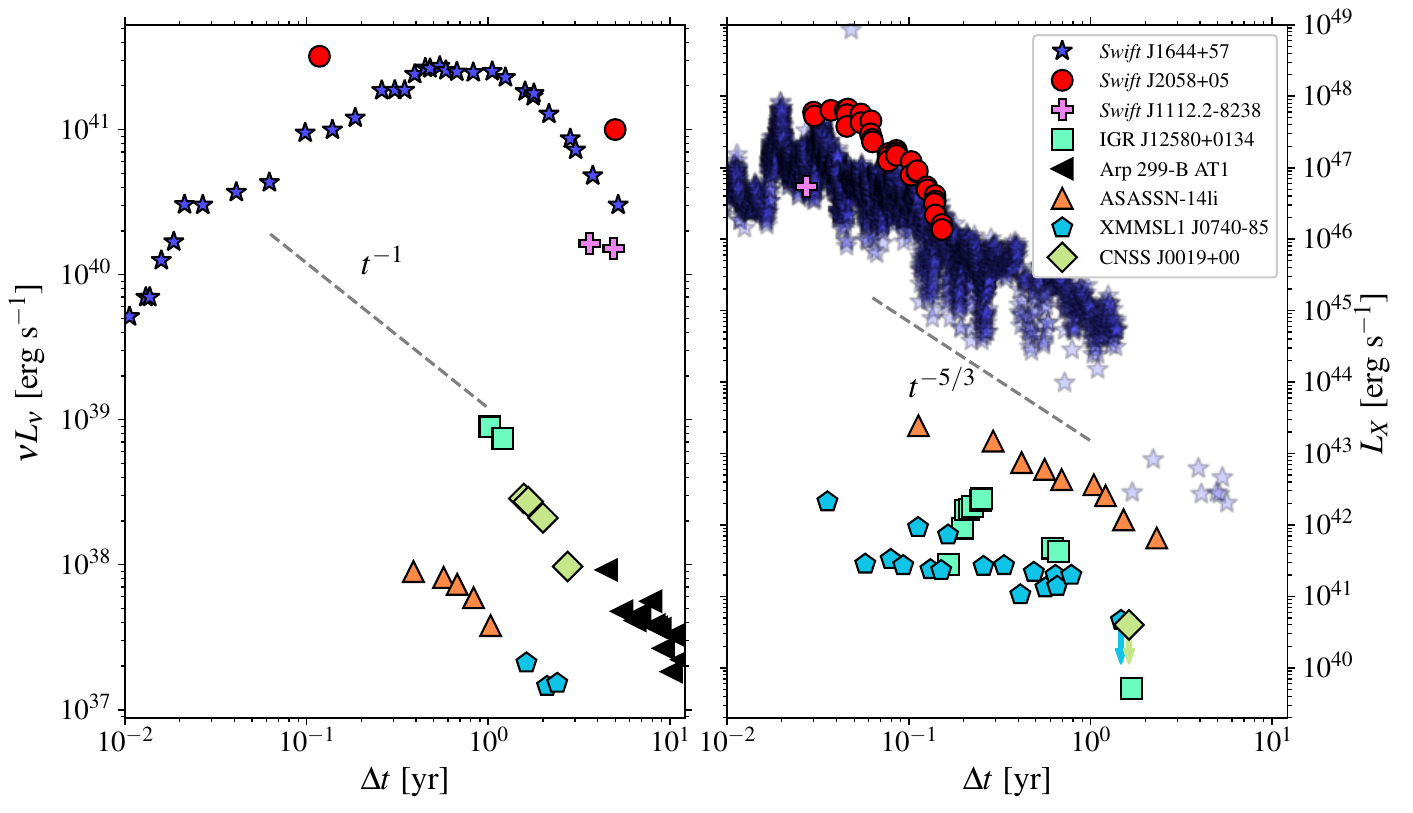}
	\caption{Radio luminosity (left) and X-ray luminosity (right) as a function of approximate time since explosion date, for all TDEs with confirmed radio detections. The reported radio luminosities correspond to frequencies between 4.5 and 6.0\,GHz. This frequency range was chosen because it samples the optically thin side of the synchrotron spectrum at late times for the thermal TDE events plotted here. The dashed line shows the expected $\nu L_\nu \propto t^{-1}$ relation for an adiabatically expanding source in the optically thin regime. The radio data are taken from the following -- for \textit{Swift} J1644+57, \citealt{Berger+2012}, \citealt{Zauderer+2013}, and \citealt{Eftekhari+2018}; for \textit{Swift} J2058+05, \citealt{Cenko+2012} and \citealt{Brown+2017b}; for \textit{Swift} J1112.2-8238, \citealt{Brown+2017b}; for IGR J12580+0134, \citealt{Irwin+2015}; for Arp 299B-AT1, \citealt{mattila2018}; for ASASSN-14li, \citealt{Alexander+2016}; for XMMSL1 J0740-85, \citealt{Alexander+2017}. The X-ray data are taken from the following -- for \textit{Swift} J1644+57, \citealt{Mangano+2016} and \citealt{Eftekhari+2018}; for \textit{Swift} J2058+05, \citealt{Cenko+2012}; for \textit{Swift} J1112.2-8238, \citealt{Brown+2015}; for IGR J12580+0134, \citealt{Lei+2016} and \citealt{Nikolajuk+Walter2013}; for ASASSN-14li, \citealt{Brown+2017}, \citealt{Holoien+2016}, and \citealt{Bright+2018}; for XMMSL1 J0740-85, \citealt{Saxton+2017}. The values with arrows are upper limits on the X-ray flux. The dashed line shows the theoretical $L_X\propto t^{-5/3}$ relation for fall-back accretion. We note that TDEs frequently deviate from this relation, but we plot it here as a general guide.}
	\label{fig:radioxrayluminosity}
\end{figure}

We fit the spectra of CNSS J0019+00 for a single spectral break, and assume that the peak frequency corresponds to the synchrotron self-absorption frequency $\nu_\text{sa}$. From Equation 1 of \citealt{Granot+Sari2002}, 
\begin{equation}\label{eq:granot+sari}
F_\nu = F_{\nu_b,\rm{ext}} \left[ \left( \frac{\nu}{\nu_b} \right)^{-s \beta_1} + \left( \frac{\nu}{\nu_b} \right)^{-s \beta_2} \right]^{-1/s},
\end{equation}
\noindent where $\nu_b$ is the frequency of the spectral break, and $F_{\nu_b,\rm{ext}}$ is the extrapolated flux density at the location of $\nu_b$ where the power laws on either side of the spectral break, $\beta_1$ and $\beta_2$, meet. The parameter $s$ describes the sharpness of the spectral break. $\beta_1$ is not well constrained from our radio follow-up observations due to the lack of data points at frequencies below 1 GHz. However, the optically thick portions of the spectra are shallower than the expected $F_\nu \propto \nu^{5/2}$ for a synchrotron self-absorbed spectrum, assuming $\nu_m < \nu_\text{sa}$, where $\nu_m$ is the frequency of the lowest energy electrons in the electron power-law distribution~\citep[see discussion of spectral breaks in][]{Granot+Sari2002}. In addition, the spectrum obtained on 2015 May 10 is consistent with $\beta_1 = 2$. This value for $\beta_1$ and the broadness of the peaks in the follow-up SEDs indicates CNSS J0019+00 was in the regime of $\nu_\text{sa} < \nu_\text{m}$, with the two break frequencies in close proximity, throughout these observations, similar to what was observed in the radio SED of \textit{Swift} J1644+57 at comparable timescales~\citep{Eftekhari+2018}. We therefore proceed with $\beta_1 = 2$. The spectral index values for the optically thin regime range between $-1$ and $-1.5$ between the five radio follow-up epochs, and are all consistent with $\beta_2 = -1.15$, within uncertainties.
We use this value of $\beta_2$ in the analysis that follows. We determine the remaining best-fit parameters from the model described by Equation~\ref{eq:granot+sari} for each of the follow-up spectra using the Python Markov Chain Monte Carlo module {\tt emcee} \citep{Foreman-Mackey+2013}. Figure~\ref{fig:vlafollowup} shows the fit to each follow-up spectrum. The best-fit values for $s$, and the peak frequency and flux from the model fit, are listed in Table~\ref{tab:bestfitparams}. Both the peak flux and the peak frequency are declining with time -- on 2015 May 10 the spectrum peaks at 8.3\,mJy at 4.3\,GHz; by 2016 July 8 the flux has declined to 5\,mJy at 1.9\,GHz. The optically thin spectral slope of $\nu^{-1.15}$ implies $p=3.3$. 

Knowledge of the synchrotron self-absorption frequency and peak flux at multiple epochs is sufficient for placing a strong constraint on the source size, or radius $R$, as a function of time, and therefore providing an estimate for the average velocity of the outflow. This estimate on the outflow velocity allows the source size to be extrapolated back in time and therefore providing an approximate date for the launch of the outflow $t_0$. As is frequently done for synchrotron-emitting systems for which the synchrotron self-Compton peak frequency and flux is not known, the assumption of equipartition~\citep{Pacholczyk1970, Scott+Readhead1977} between the energy in electrons and the energy in the magnetic field allows for the calculation of the equipartition size (or $R_\text{eq}$ for an assumed source geometry) and the minimum total equipartition energy, $E_\text{eq}$, as well as the magnetic field strength and local density~\citep[see, e.g.,][]{Chevalier1998}. The equations used to compute these micro and macrophysical parameters for CNSS\,J0019+00 are detailed in Appendix~\ref{ap1}. The values determined from the best-fit model to the radio SEDs are given in Table~\ref{tab:bestfitparams}.

Figure~\ref{fig:equipartition} shows the values of $R_\text{eq}$ and $E_\text{eq}$ as a function of time, for each of our radio follow-up observations. The source nearly doubles in size on the 14 month timescale probed by our first and fourth follow-up epochs on 2015 May and 2016 July, from an equipartition radius of $R_\text{eq} \sim 7 \times 10^{16}\,\text{cm}$ to $13 \times 10^{16}\,\text{cm}$. This corresponds to an average expansion velocity of $v_\text{ej} \approx 15,000\,\text{km\,s}^{-1}$, with the equipartition radii well fit by a constant expansion velocity. From this we extrapolate the outflow back in time, to determine the age of the event at each of our observation epochs ($\Delta t \approx R_\text{eq} / v_\text{ej}$). The outflow was launched on 2013 October 15, 522 days prior to the CNSS epoch 4 observation in which it was discovered on 2015 March 21.

\begin{figure}[htb!]
\centering
	\subfigure[]{\includegraphics[width=0.49\textwidth]{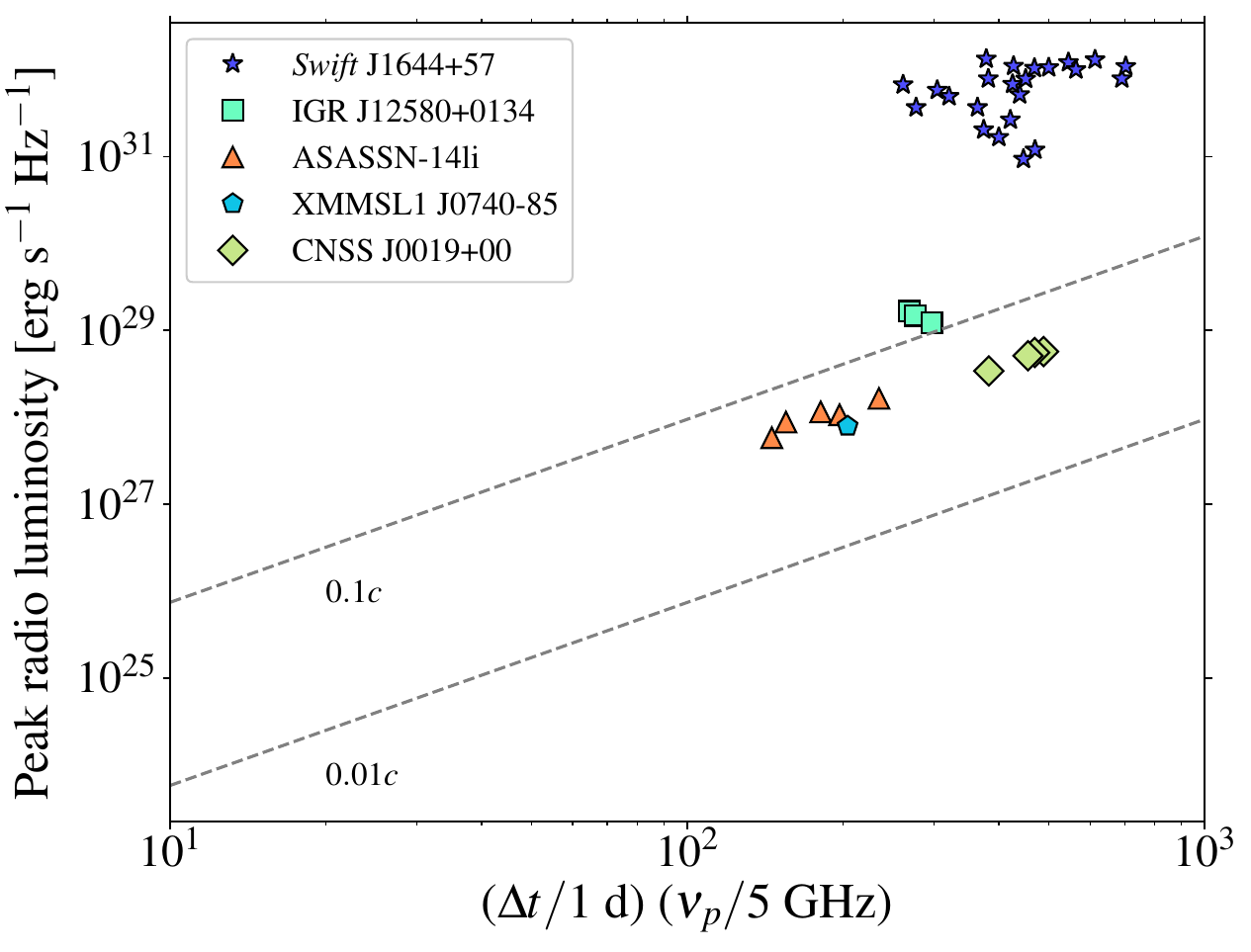}}
	\subfigure[]{\includegraphics[width=0.49\textwidth]{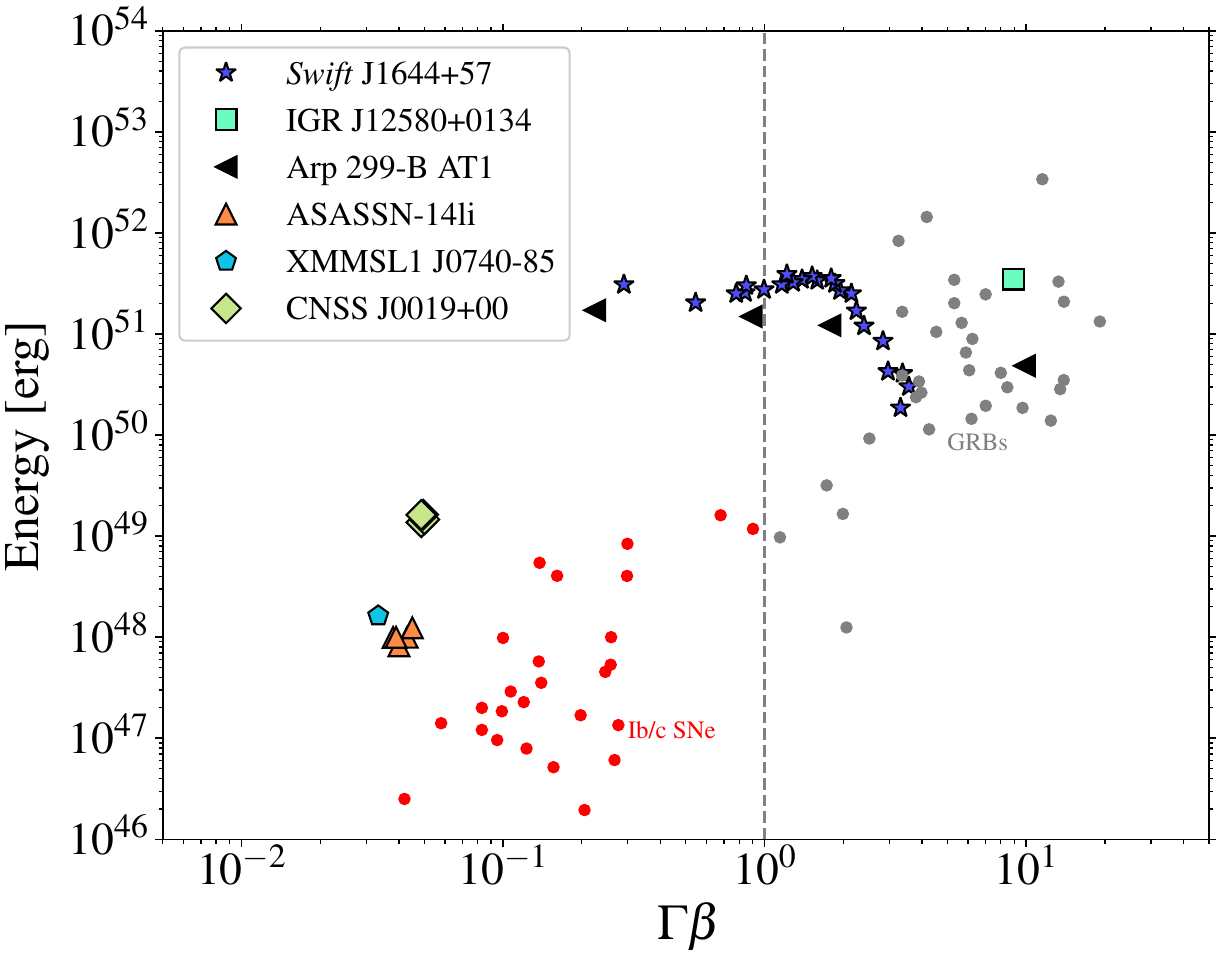}}
	\caption{Peak radio luminosity as a function of the product of the peak time and peak frequency (left). The outflow velocity is proportional to $F_{p,\text{mJy}}^{(6+p)/(13+2p)} \, \Delta t^{-1} \, \nu_p^{-1}$, and can therefore be represented on the plot with lines of constant velocity (assuming a constant $p$; see Figure 4 of \citealt{Chevalier1998}). The dashed lines here are consistent with our method and set of assumptions as outlined in the Appendix~\ref{ap1}, and using a value of $p=3.3$. The values for the radio luminosity, frequency, and time of peak for each object are determined from model fits to the data, rather than directly from the observations due to the fact that the radio SEDs are frequently poorly sampled. We also plot the total energy as a function of outflow velocity (right). The area to the left (right) of the dashed line represents non-relativistic (relativistic) outflows. CNSS J0019+00 has a constant expansion velocity that is consistent with the other radio-detected thermal TDEs (equipartition energies for these objects were recomputed to be consistent with the method and set of assumptions use in this paper), however it is more than an order of magnitude more energetic. The values from the model fits are taken from -- for \textit{Swift} J1644+57, \citealt{Eftekhari+2018}; for IGR J12580+0134, \citealt{Lei+2016}; for Arp 299B-AT1 and the sample of GRBs and Ib/c SNe, \citealt{mattila2018} and references therein; for ASASSN-14li, \citealt{Alexander+2016}; for XMMSL1 J0740-85, \citealt{Alexander+2017}.}
	\label{fig:LandEcomparison}
\end{figure}

Figures~\ref{fig:radioxrayluminosity} and~\ref{fig:LandEcomparison} place CNSS J0019+00 in the context of other radio-detected TDEs. There is a clear divide between the non-thermal, jetted TDEs\footnote{IGR J12580+0134 and Arp 299-B AT1 are under-luminous relative to their non-thermal, jetted counterparts but are likely non-thermal events for which the jet was launched off-axis.} like \textit{Swift} J1644+57, and the thermal, non-jetted TDEs like ASASSN-14li --- both in their respective radio and X-ray luminosities (Figure~\ref{fig:radioxrayluminosity}), as well as in velocity-energy space (Figure~\ref{fig:LandEcomparison}). CNSS J0019+00's radio luminosity and outflow velocity are consistent with the thermal, non-jetted TDEs. We note that our upper limit on the X-ray luminosity of CNSS J0019+00 is comparable to that of XMMSL1 J0740-85 at a similar post-explosion date. It is also consistent with the luminosities, both measurements and upper limits, of the larger sample of thermal X-ray TDEs~\citep{Auchettl+2017}. This division between the majority of TDEs with non-relativistic outflows and the $\sim$few percent of more energetic TDEs that launch relativistic jets reflects a similar division present between Type Ib/c SNe and long GRBs (Figure~\ref{fig:LandEcomparison}).

\section{Host Galaxy, SDSS J0019+00}\label{sec:host}

\begin{figure}[htb!]
\centering
    \raisebox{0.2in}{\includegraphics[width=0.18\textwidth]{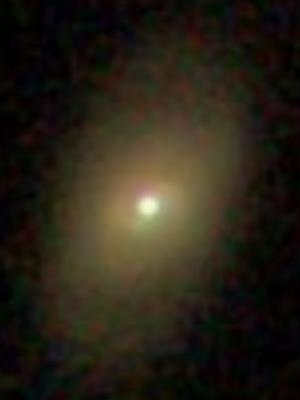}\hspace{0.1in}}
	\includegraphics[width=0.38\textwidth]{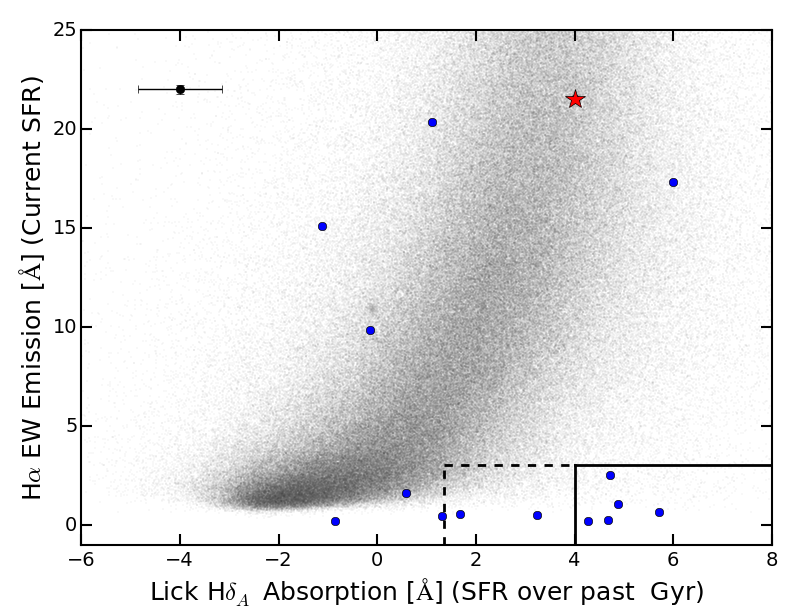}
	\includegraphics[width=0.38\textwidth]{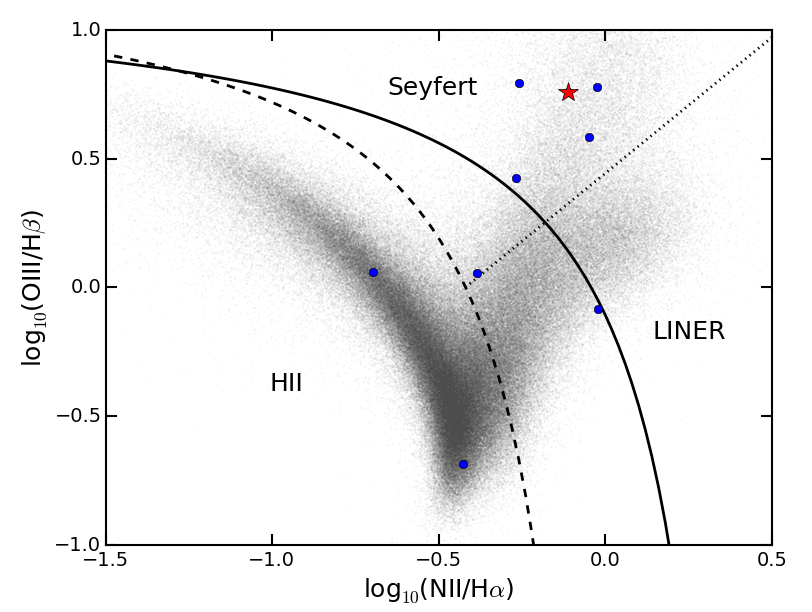}
	\caption{CNSS J0019+00 host galaxy (SDSS J0019+00) properties. {Left:} Three-color SDSS image cutout ($30\arcsec\times40\arcsec$) of the host galaxy of CNSS J0019+00, located at $z=0.018$ (77 Mpc). 
 	{Middle:} Plot of the H$\alpha$ emission line equivalent widths versus the Lick H$\delta_A$ indices. These parameters measure the current and past star formation rates respectively. SDSS galaxies (MPI-JHU catalog; https://wwwmpa.mpa-garching.mpg.de/SDSS/DR7/) are shown in gray, optically-selected TDE hosts \citep[sample from][]{Law-Smith+2017,french2016,french2017,mattila2018,holoien2019} are shown as blue circles, and the host galaxy of CNSS J0019+00 is shown as a red star. The H$\alpha$ emission from SDSS J0019+00 is dominated by the AGN, and the shows enhanced Lick H$\delta_A$ index, similar to the optically-selected TDE hosts. The solid and dashed lines in the bottom-left corner demarcate the region corresponding to 0.2\% of SDSS galaxies and 2\% of SDSS galaxies respectively. The errorbar on the top right represents the typical uncertainty in the Lick H$\delta_A$ index. {Right:} The BPT (O\,III/H$\beta$ versus N\,II/H$\alpha$) diagram. SDSS galaxies, optically-selected TDE hosts, and the  host galaxy of CNSS J0019+00 are shown with similar markers as the middle panel. SDSS J0019+00 lies securely in the AGN/Seyfert region of the BPT diagram. See Section \ref{sec:host} for details.}
	\label{fig:host}
\end{figure}

The three-color image cutout of host galaxy of CNSS J10019+00 is shown in the left panel of Figure~\ref{fig:host}. 
SDSS J0019+00 is a S0 galaxy \citep{huertas-company2011}, lacking any evidence of large-scale spiral arms, having a bright nucleus reminiscent of Seyfert 2 galaxies.
The yellow/red appearance of the galaxy in the three-color image indicates that there is no substantial ongoing star formation.
A barred spiral structure, having a angular scale of 5$\arcsec$--10$\arcsec$\,(diameter), is seen in the three-color image.
There may be a fainter disk/shell-like feature extending beyond this spiral structure, possibly indicative of a past merger, although this will have to be verified through a more substantive analysis.
WISE colors, $W2-W3\simeq2.8$ and $W1-W2\simeq0$, are consistent with spirals/LIRGs \citep{wright2010}.
The SDSS light curve between 2002 and 2008 shows small amplitude optical variability (about 0.5 mag in \textit{u} and \textit{r} bands), indicating low-level (optical) AGN activity.
Absence of radio detection of SDSS J0019+00 in the first three epochs of the CNSS, as well as any archival radio data,  suggests the absence of a persistent radio jet associated with the central SMBH. 

The stellar velocity dispersion from SDSS spectroscopy is $\sigma\simeq70$ km\,s$^{-1}$ \citep[e.g.][]{nair2010}, and using this in the $M_{\text{BH}}$--$\sigma$ relationship \citep{mcconnell_ma2013} we can estimate the black hole mass\footnote{We note that the measured value of 70 km\,s$^{-1}$ is below the resolution limit of the SDSS spectrograph, and this estimate should be treated as very approximate or a rough upper limit on the black hole mass.} to be $M_{\text{BH}}\simeq 10^6 \text{M}_\sun$.
Based on the 4000$\text{\normalfont\AA}$ break strength and the Balmer absorption-line index $H\delta_A$, \cite{kauffmann2003} estimate the mean stellar age of SDSS J0019+00 to be about 1.5 Gyr.
The specific star formation rate (sSFR) is about $5\times10^{-11}$ per year \citep[e.g.][]{chang2015,vanDriel2016}. These findings confirm that SDSS J0019+00 consists primarily of old stars, and lacks significant ongoing star formation.

Bulge-disk decomposition of SDSS J0019+00 has been carried out by \cite{simard2011}\footnote{The disk inclination is found to be approximately $50^\circ$ \citep{simard2011}.}. 
The bulge/total surface brightness ratio is 0.59 and Sersic index is 6.2.
The bulge semi-major effective radius is 0.66 kpc (bulge ellipticity is equal to the galaxy ellipticity, 0.3).
This can be compared to the \textit{g}-band galaxy semi-major axis (half-light radius), 2.17 kpc.
\cite{mendel2014} find that the bulge stellar mass is about $3\times10^9$\,$M_\sun$, compared to a total stellar mass of $9\times10^9$\,$M_\sun$. All of these measurements indicate a high central concentration of stars, consistent with the trend of TDE host galaxies having more centrally concentrated stellar populations~\citep{Law-Smith+2017, french2020}.

The middle panel of Figure~\ref{fig:host} plots the H$\alpha$ emission line equivalent widths versus the Lick H$\delta_A$ indices (or equivalently, the current versus past star formation rates) for SDSS galaxies, optically-selected TDE host galaxies and SDSS J0019+00. The enhanced Lick H$\delta_A$ index indicates that SDSS J0019+00 is very similar to other TDE hosts in terms of stellar composition (abundance of A-type stars). The H$\alpha$ emission from SDSS J0019+00 is dominated by the AGN, and hence the apparent distinct location of this galaxy, compared to other TDE hosts, on this plot.
SDSS J0019+00 could be related to shocked post-starburst galaxies, e.g. the host galaxy of ASASSN-19bt \citep[][see Figure 2 of that paper]{holoien2019}, which are similar to E+A galaxies that are known to host several optically-selected TDEs but have higher dust obscuration.
The right panel of Figure~\ref{fig:host} shows the BPT diagram for SDSS J0019+00, which shows the galaxy being located distinctly above the canonical line separating star-forming galaxies from AGNs. Nebular line flux ratios from the SDSS spectrum (Figure~\ref{fig:host}) indicate that this galaxy is consistent with being a Seyfert 2.

Taken together, the galaxy morphology and old stellar population point toward SDSS J0019+00 being consistent with a typical TDE host galaxy. We will return to this point in the next section.


\section{Summary \& Discussion}\label{discussion}
We have reported on the discovery of the radio transient CNSS J0019+00, which was found during CNSS, a dedicated transient survey carried out with the VLA.
Triggered radio follow-up observations together with our equipartition analysis suggests a $\sim15,000$\,km\,s$^{-1}$ outflow having energy of approximately $10^{49}\,\text{erg}$. We note this is consistent with the predictions of~\citealt{Lu+Bonnerot2019} for a non-relativistic collision-induced outflow in TDEs. The transient is located on the nucleus of a Seyfert 2 galaxy at a redshift of $z=0.018$.
The position of the galaxy nucleus in \textit{Gaia} and the location of the radio transient in our VLBA observations are consistent to within 1\,pc. Taken together, this indicates that CNSS J0019+00 is likely the first radio-discovered TDE, and possibly the third such radio detection of a non-jetted TDE.

We now consider possible alternative explanations for CNSS J0019+00.
Type II supernovae (SNII) are among the class of radio transients that have the largest rates \citep{gal-yam2006,Mooley+2016} and have spectral evolution similar to that observed for our CNSS transient.
Hence we explore the possibility of CNSS J0019+00 being a SNII.
Firstly, we find that the host galaxy, SDSS J0019+00, is unusual for SNII: it is an S0 galaxy, has low sSFR, relatively high stellar mass, and no evidence for recent star formation (i.e. relatively old stellar population; see \S\ref{sec:host}).
The SNII volumetric rate for S0 galaxies (about $5\times10^{-8}$ Mpc$^{-3}$ yr$^{-1}$; \citet{LOSSIII}) is more than an order of magnitude lower than that for late-type galaxies. Second, given the stellar mass distribution within the host galaxy (see \S\ref{sec:host}), we expect the probability of finding an SN in the non-nuclear regions of the host to be somewhat larger than that within the nuclear region. 
Thirdly, finding a SNII very close to the central SMBH (CNSS J0019+00 is consistent with the nucleus to within 1 pc) is unlikely.
Even in extreme cases like the nuclear supernova ``factory" Arp 299  \citep{neff2004,perez-torres2010}, where the spatial density of SNe is high, the probability of finding a supernova within 1 pc of the nucleus is $<$1\%.
Taken together, we find that the probability of CNSS J0019+00 being a nuclear supernova is minuscule.
We note also that the peak radio luminosity of CNSS J0019+00, $5\times10^{28}$ erg s$^{-1}$ Hz$^{-1}$ at 3 GHz, is at the tail end of the luminosity distribution observed for optically-selected SNII \citep[e.g.][]{weiler2002}, but not extremely unusual for radio-selected SNe.

We also consider the possibility of CNSS J0019+00 being renewed jet activity from an AGN \citep{Mooley+2016}.
Renewed jet activity (possibly due to a sudden gas accretion event) has been inferred in some radio AGNs over a $\sim$decade timescale from the CNSS Pilot survey \citep{Mooley+2016}, where the new radio sources associated with these AGNs have long-lasting ($>$5 years) radio emission with luminosities of $\gtrsim$10$^{29}$ erg s$^{-1}$ Hz$^{-1}$.
We find that the timescale ($\sim$2 yr at frequencies of a few GHz) and energetics of CNSS J0019+00 are not consistent with this AGN population.
Nevertheless, we cannot conclusively rule out a renewed AGN jet hypothesis for CNSS J0019+00; if this radio transient is indeed such an AGN event, then we conclude that the jet is not long-lasting (like the events found in the CNSS Pilot survey), but becomes luminous and fades away over a timescale of 1--2 yr.

The rates, timescales, and environments for other kinds of extragalactic transient events, such as off-axis GRBs, are very different from those of CNSS J0019+00 (see below; also \cite{Mooley+2016} and references therein).  
We therefore conclude, based on the host galaxy properties (which are similar to other TDE hosts), spectral evolution, association with the nucleus of its host galaxy, and transient rate, that CNSS J0019+00 is a TDE.

Given that TDE is the likely explanation for CNSS J0019+00, we derive constraints on the mass of the disrupted star. From our equipartition analysis we find that the outflow has not decelerated, so we can calculate a lower limit on the ejecta mass, $M_{\rm ej} \gtrsim 5\times10^{-3}$\,$M_\odot$.
The mass of the star is therefore $\gtrsim$0.1\,$M_\odot$ ($f_{\rm ej}$/0.1)$^{-1}$ ($\eta$/0.5)$^{-1}$, where $f_{\rm ej}$ is the fraction of the stellar mass that goes into the radio-emitting ejecta, and $\eta$ ($\leqslant$1) is a fudge factor that is influenced by ejecta velocity stratification, radiative efficiency etc.

CNSS J0019+00 presents an excellent opportunity to measure the density profile of material around the SMBH of the host galaxy. 
Assuming that the microphysical parameters ($\epsilon_B$, $\epsilon_e$, $\gamma_m$) have remained unchanged throughout the propagation of the blastwave through the circumnuclear environment, we can work out the radial profile of the electron density, $n\propto r^{-2.5}$ between 5--15$\times$10$^{16}$\,cm (about 2--5$\times$10$^{5}$ $R_S$) from the central SMBH of SDSS J0019+00 (see \S\ref{analysis}). 
This profile is similar to the one deduced for ASASSN-14li \citep{Alexander+2016} between  1--4$\times$10$^{16}$\,cm (about 0.5--2$\times$10$^{5}$ $R_S$) from the nucleus of the host galaxy.
For radiatively inefficient accretion flows (RIAFs) we expect a gas density profile $n(r) \propto r^{-\gamma}$ where $0.5<\gamma<1.5$ ($\gamma=1.5$ is typical for Bondi/advection-dominated  accretion; e.g. \cite{quataert_gruzinov2000}).
For example, in the case of Sgr A* $\gamma\simeq1$ within $\sim$10$^3$ $R_S$ and $\gamma\simeq0.5$ beyond this distance \citep[][and references therein; see also Figure~\ref{fig:density}]{gillessen2019}.
In the case M87, \cite{russell2018} find $\gamma\simeq0.9-1.5$ within the Bondi radius, implying inflow perpendicular to the jet axis and an outflow along the jet axis.
The density profile in the case of the jetted TDE Swift J1644+57 \citep[e.g.][]{Eftekhari+2018} is consistent with $\gamma\simeq1-1.5$ at about 10$^6$ $R_S$, with some indication of steepening above and below this radius.
The steep density profile seen in the nucleus of SDSS J0019+00 implies an accretion flow that is quite different from RIAF, and may indicate a substantial rate for the outflow of material from the central SMBH. In all cases, the (extrapolated) density lies between 10$^3$--10$^5$ cm$^{-3}$ at the Bondi radius ($\sim$10$^5$ $R_S$) and consistent with 1 cm$^{-3}$ at $\sim$few$\times$10$^6$ $R_S$.

Using the CNSS survey we can calculate, for the first time, an unbiased rate of TDE outflows similar to CNSS J0019+00. 
Given a timescale of a few months for the spectral evolution at 3 GHz, we have only two effective epochs of observing (each of which was over 270 deg$^2$ of the Stripe 82 region, and have a source detection limit of about 500 $\mu$Jy) within the CNSS survey.
We therefore calculate the TDE outflow rate to be $R(>500 \mu{\rm Jy})=1.8^{+5.4}_{-1.6}\times10^{-3}$\,deg$^{-2}$ of the sky (90\% confidence interval, assuming Poisson statistics; \citealt{gehrels1986}). Alternatively, we can find the volume, corresponding to the peak radio luminosity of CNSS J0019+00, accessible to the CNSS and calculate a volumetric rate of about 10 Gpc$^{-3}$ yr$^{-1}$. This is an order of magnitude larger than the expected rate of jetted TDEs (\textit{Swift} J1644+57-like events that are seen off-axis, assuming a beaming fraction of 100), $\sim$1 Gpc$^{-3}$ yr$^{-1}$ \citep[e.g.][]{metzger2015}.
Thus, observationally we find that Newtonian outflows accompanying TDEs are much more ubiquitous than jets in TDEs. 
We can also compare the rate of radio-selected TDEs (CNSS 0019+00-like events) with that of optically-selected TDEs, $\sim$50 Gpc$^{-3}$ yr$^{-1}$ \citep{vanVelzen+Farrar2014}. The radio TDE rate therefore represents $\sim$20\% of the rate of optically-selected TDEs, despite probing the same approximate volume (within $z=0.1$), though we note that this discrepancy is not unexpected given the uncertainty in the radio luminosity function of thermal TDEs\footnote{It is possible that a significant number of radio-selected TDEs may be dust-obscured and therefore invisible at optical/UV wavelengths.}~\citep{Alexander+2020, Stone+2020}. Finally, we note that, given our rate of radio TDEs, we expect to find tens of events like CNSS J0019+00 in all-sky radio surveys being executed with the VLA (VLASS; \citealt{lacy2019}) and ASKAP \citep{murphy2013} --- more numerous than the number of TDEs expected previously.

\acknowledgements 
Acknowledgements: The authors thank Ehud Nakar for insightful discussion on equipartition analysis, Vikram Ravi for helpful comments on improving this paper, Yi Cao for the optical data reduction analysis, Mansi Kasliwal for the use of CLU in the identification of transient events in CNSS. The authors would also like to thank the anonymous referee for valuable feedback in improving the quality of this paper. KPM is a Jansky Fellow of the National Radio Astronomy Observatory.
The National Radio Astronomy Observatory is a facility of the National Science Foundation operated under cooperative agreement by Associated Universities, Inc. AH acknowledges support by the I-Core Program of the Planning and Budgeting Committee and the Israel Science Foundation. This research was supported by grant No.\,2018154 from the United States-Israel Binational Science Foundation (BSF).

\appendix
\section{Equipartition Analysis for a Synchrotron-emitting System}\label{ap1}
In solving for the equipartition parameters for a synchrotron-emitting source, we follow the steps of~\citealt{Chevalier1998}, for a spherical, non-relativistically expanding source. For a synchrotron self-absorbed system, the minimal equipartition energy $E_\text{eq}$, radius $R_\text{eq}$, and magnetic field $B_\text{eq}$ can be determined from the peak flux $F_{\nu_p}$ and frequency $\nu_p$ at which the spectrum transitions from optically thick to optically thin. These will also be dependent on the distance to the source $D$ and the power law index of the electron energy distribution $p$, where $N_e(E) = N_o\, E^{-p}$ is the density of relativistic electrons per unit energy, where $p$ is determined from the slope of the optically thin side of the spectrum. There is only a very weak dependence of the microphysical parameters on the value of $p$, however we include it here for completeness.

Under the assumption that the observed peak in the synchrotron spectrum is due to self-absorption, we can write the flux in the optically thick and thin limits as $F_p$. In the optically thick limit,
\begin{equation}\label{eq:oldthick}
F_\nu = \frac{\pi R^2}{D^2} \, \frac{j_\nu}{\alpha_\nu},
\end{equation}
and in the optically thin limit,
\begin{equation}\label{eq:oldthin}
F_\nu = 4 \pi j_\nu \, \frac{4}{3}\pi R^3 f \, \frac{1}{4 \pi D^2},
\end{equation}
where $f$ is the emission filling factor for a spherical emission region with outer radius $R$, $\alpha_\nu$ is the synchrotron absorption coefficient, and $j_\nu$ is the synchrotron emission coefficient. From~\citealt{Rybicki+Lightman1979}, $\alpha_\nu$ is given by
\begin{equation}
\alpha_\nu = c_6 N_o B^{(p+2)/2} \left( \frac{\nu}{2c_1} \right)^{-(p+4)/2},
\end{equation}
where the constants $c_6$ and $c_1$ are given by
\begin{equation*}
2 c_1 = \frac{3e}{2 \pi m_e^3 c^5}
\end{equation*}
\begin{equation*}
c_6 = \frac{\sqrt{3} e^3}{8 \pi m_e} \, \left( \frac{3 e}{2 \pi m_e^3 c^5} \right)^{-2} \, \Gamma\left( \frac{3p+2}{12} \right) \, \Gamma\left( \frac{3p+22}{12} \right).
\end{equation*}

From~\citealt{Rybicki+Lightman1979}, $j_\nu$ is given by\footnote{Derived here in terms of the electron energies $E$, rather than $\gamma_e$ as in \citealt{Rybicki+Lightman1979}.}
\begin{equation}
j_\nu = c_5 N_o B^{(p+1)/2} \left( \frac{\nu}{2c_1} \right)^{-(p-1)/2},
\end{equation}
where the constant $c_5$ is given by\footnote{$c_5$ and $c_6$ are the constants tabulated as a function of $p$ by~\citealt{Pacholczyk1970}.}
\begin{equation*}
c_5 = \frac{\sqrt{3} e^3}{4 \pi m_e c^2 (p+1)} \, \Gamma\left(\frac{p}{4} + \frac{19}{12}\right) \, \Gamma\left( \frac{p}{4} - \frac{1}{12} \right).
\end{equation*}

We can now rewrite Equations~\ref{eq:oldthick} and~\ref{eq:oldthin} as
\begin{equation}\label{eq:thick}
F_\nu = \frac{\pi R^2}{D^2} \frac{c_5}{c_6} B^{-1/2} \left(\frac{\nu}{2 c_1} \right)^{5/2}
\end{equation}
\begin{equation}\label{eq:thin}
F_\nu = \frac{4 \pi R^3 f}{3 D^2} c_5 N_o B^{(p+1)/2} \left( \frac{\nu}{2 c_1} \right)^{-(p-1)/2}.
\end{equation}

The constant $N_o$ is determined by the equipartition analysis between the energy density in relativistic electrons $u_e$ and the energy density in the magnetic field $u_B$. We follow the convention of~\citealt{Chevalier1998} and use the electron rest mass energy $E_l = 0.51$\,MeV as the lower bound of the relativistic electron energy density distribution. Then,
\begin{equation*}
\int_{E_l}^{\infty} N(E) E dE = \frac{u_e}{\epsilon_e} = \frac{u_B}{\epsilon_B},
\end{equation*}
where $\epsilon_e/\epsilon_B$ is the ratio of relativistic electron energy density to magnetic energy density. Then,
\begin{equation}\label{eq:densityconst}
N_o = \left( \frac{\epsilon_e}{\epsilon_B} \right) \frac{B^2}{8 \pi} (p-2) E_l^{p-2}.
\end{equation}

Combining Equations~\ref{eq:thick},~\ref{eq:thin}, and~\ref{eq:densityconst}, and evaluating the flux and frequency at the peak of the spectrum as $F_p$ and $\nu_p$, we can solve for the equipartition radius and magnetic field:
\begin{equation}
R_\text{eq} = \left[ \frac{6 c_6^{p+5} F_p^{p+6} D^{2p+12}}{(\epsilon_e/\epsilon_B) f (p-2) \pi^{p+5} c_5^{p+6} E_l^{p-2}} \right]^{1/(2p+13)} \, \left( \frac{\nu_p}{2c_1} \right)^{-1}
\end{equation}
\begin{equation}
B_\text{eq} = \left[ \frac{36 \pi^3 c_5}{(\epsilon_e/\epsilon_B)^2 f^2 (p-2)^2 c_6^3 E_l^{2(p-2)} F_p D^2} \right]^{2/(2p+13)} \, \left( \frac{\nu_p}{2c_1} \right).
\end{equation}
These are Equations 11 and 12 of~\citealt{Chevalier1998}. The equipartition energy is given by
\begin{equation}
E_\text{eq} = \frac{B_\text{eq}^2}{\epsilon_B 8 \pi} \, \frac{4}{3} \pi R_\text{eq}^3 f.
\end{equation}

The density is given by
\begin{equation}
n = \int_{E_l}^{\infty} N_o E^{-p} dE = \left( \frac{\epsilon_e}{\epsilon_B} \right) \frac{B_\text{eq}^2}{8 \pi} \frac{(p-2)}{(p-1)} E_l^{-1}.
\end{equation}

Because the slope of the optically thin spectra for CNSS\,J0019+00 is consistent with $p=3.3$, we use the following values for the constants: $c_1 = 6.27 \times 10^{18} \,\text{g}^{-5/2}\,\text{cm}^{-7/2}\,\text{s}^{4}$, $c_5 = 6.68 \times 10^{-24}\,\text{g}^{1/2}\,\text{cm}^{5/2}\,\text{s}^{-1}$, and $c_6 = 8.08 \times 10^{-41}\,\text{g}^{11/2}\,\text{cm}^{23/2}\,\text{s}^{-11}$.

The equations above were derived under the assumption that $\nu_p = \nu_\text{sa}$. Although the radio SEDs of CNSS J0019+00 are in the regime of $\nu_\text{sa} < \nu_m$, because the break frequencies are close, the correction to the above equations is small and not considered here.

For CNSS J0019+00, we use the simplifying assumption that the energy is equally divided between the electrons, protons, and magnetic fields ($\epsilon_e = \epsilon_B = 1/3$). However, in practice, the range of possible values for these parameters spans orders of magnitude, and the choice of parameters has a significant impact on the resulting equipartition energy (see e.g.,~\citealt{Ho+2019} Section 4.1 for a discussion of conventions and assumptions regarding $\epsilon_e$ and $\epsilon_B$). For \textit{Swift} J1644+57, \citealt{Eftekhari+2018} directly constrain these parameters through observations of the synchrotron cooling break, to $\epsilon_e = 0.1$ and $\epsilon_B = 10^{-3}$, which is far below equipartition. This choice of parameters for CNSS J0019+00 increases the equipartition energies by more than an order of magnitude. However, because $\epsilon_e$ and $\epsilon_B$ are unconstrained in the case of CNSS J0019+00, we proceed with the simplifying assumption above.

\bibliography{references}

\end{document}